%% file: ms.tex
\documentclass[11pt]{article}
\usepackage{fullpage}
\usepackage{enumerate}
\usepackage{url}
\usepackage{latexsym}
\usepackage{amsmath}
\usepackage{csvsimple}
\usepackage{graphicx}
\usepackage{CJKutf8}

\usepackage[colorinlistoftodos]{todonotes}
\newtheorem{theorem}{Theorem}
\newtheorem{lemma}[theorem]{Lemma}

\newtheorem{proposition}[theorem]{Proposition}
\newtheorem{remark}[theorem]{Remark}

\newenvironment{proof}{\paragraph{Proof.}}{\hfill$\Box$}

\begin{document}


%
%

\title{m-Bonsai: a Practical Compact Dynamic Trie\thanks{This research was supported in part by the Academy of Finland via grant 294143.}}

\newcommand{\get}{\mbox{\textit{get}}} 
\newcommand{\set}{\mbox{\textit{set}}} 
\newcommand{\create}{\mbox{\textit{create}}} 

\newcommand{\Sum}{\mbox{\textit{sum}}} 
\newcommand{\update}{\mbox{\textit{update}}} 
\newcommand{\search}{\mbox{\textit{search}}} 

\newcommand{\getChild}{\mbox{\textit{getChild}}} 
\newcommand{\addChild}{\mbox{\textit{addChild}}}
\newcommand{\delLeaf}{\mbox{\textit{delLeaf}}}
\newcommand{\addLeaf}{\mbox{\textit{addLeaf}}}
\newcommand{\getLabel}{\mbox{\textit{getLabel}}}

\newcommand{\nextNodeLink}{\mbox{\textit{nextNodeLink}}} 
\newcommand{\getParent}{\mbox{\textit{getParent}}} 
\newcommand{\getRoot}{\mbox{\textit{getRoot}}} 

\newcommand{\project}{\mbox{\textit{project}}} 

\newcommand{\selectZero}{\mbox{\textit{select$_0$}}} 
\newcommand{\rankZero}{\mbox{\textit{rank$_0$}}} 

\newcommand{\selectOne}{\mbox{\textit{select$_1$}}} 
\newcommand{\rankOne}{\mbox{\textit{rank$_1$}}} 
\newcommand{\flip}{\mbox{\textit{flip}}}

\author{Andreas Poyias\\
Department of Informatics, University of Leicester\\
University Road, Leicester, LE1 7RH, United Kingdom.\\
\texttt{ap480@leicester.ac.uk}
\and Simon J. Puglisi\\
Helsinki Institute of Information Technology,\\
Department of Computer Science, University of Helsinki,\\
P. O. Box 68, FI-00014, Finland.\\
\texttt{puglisi@cs.helsinki.fi}
\and Rajeev Raman\\
Department of Informatics, University of Leicester\\
University Road, Leicester, LE1 7RH, United Kingdom.\\
\texttt{r.raman@leicester.ac.uk}}

\maketitle

\input{abstract.tex}


\input{intro.tex}


\input{preliminaries.tex}

\section{m-Bonsai}
\input{bonsai_new.tex}

\input{bonsai_traverse.tex}
\input{bonsai_new_conclude.tex}


\section{Implementation and experimental evaluation}

\input{experimental.tex}

\section{Conclusion}

\input{conclusion.tex}

\bibliographystyle{plain}
\bibliography{addrefs}
\end{document}

%% file: abstract.tex
\begin{abstract}
  We consider the problem of implementing a space-efficient
  \emph{dynamic trie}, with an emphasis on good practical performance.
  For a trie with $n$ nodes with an alphabet of size $\sigma$,
  the information-theoretic lower bound is $n \log \sigma + O(n)$ bits.
  The Bonsai data structure is a compact trie proposed by Darragh et al.
  (Softw., Pract. Exper. 23(3), 1993, pp. 277–291).
  Its disadvantages include the user having to specify an upper bound
  $M$ on the trie size in advance (which cannot be changed easily after 
  initalization),
  a space usage of $M \log \sigma + O(M \log \log M)$ (which
  is asymptotically non-optimal for smaller $\sigma$ or if $n \ll M$)
  and a lack of support for deletions.
  It supports traversal and update operations in $O(1/\epsilon)$ expected
  time (based on assumptions about the behaviour of hash functions), where
  $\epsilon = (M-n)/M$ and has excellent speed performance in practice.
  We propose an alternative, \emph{m-Bonsai}, that addresses the above
  problems, obtaining a trie that uses $(1+\beta) n (\log \sigma + O(1))$
  bits in expectation, and supports traversal and update
  operations in $O(1/\beta)$ expected time  and $O(1/\beta^2)$ amortized
  expected time, for any user-specified parameter $\beta > 0$ (again based on
  assumptions about the behaviour of hash functions). 
  We give an implementation of m-Bonsai which
  uses considerably less memory and is slightly faster than the original Bonsai.
\end{abstract}

%% file: intro.tex
\section{Introduction}

In this paper, we consider \emph{practical} approaches to the problem of
implementing a \emph{dynamic trie} in a highly space-efficient manner.  
A dynamic trie (also known
as a dynamic \emph{cardinal} 
tree \cite{DBLP:journals/algorithmica/BenoitDMRRR05}) 
is a rooted tree, where
each child of a node is labelled with a distinct symbol
from an alphabet $\Sigma = \{0,\ldots,\sigma-1\}$. We consider
dynamic tries that support the following operations:

\begin{description}
\item[$\create()$:] create a new empty tree.
\item[$\getRoot()$:] return the root of the current tree.
\item[$\getChild(v,c)$:] return child node of node $v$ having symbol $c$, 
if any (and return $-1$ if no such child exists).
\item[$\getParent(v)$:] return the parent of node $v$.
\item[$\getLabel(v)$:] return the label (i.e. symbol) of node $v$.
\item[$\addLeaf(v,c)$:] add a new child of $v$ with symbol $c$ and
return the newly created node.
\item[$\delLeaf(v,c)$:] delete the child of $v$ with symbol $c$, provided
that the child indicated is a leaf (if the user asks to delete a child that
is not a leaf, the subsequent operations may not execute correctly).
\end{description}
The trie is a classic data structure (the name dates back to 1959~\cite{trie1959})
that has numerous applications in string processing.  A naive implementation
of tries uses pointers.  Using this approach, each node in
an $n$-node binary trie uses $3$ pointers for the navigational operations. 
A popular alternative for larger alphabets is 
the \emph{ternary search tree (TST)} \cite{Dobbs:1998:Bentley}, 
which uses $4$ pointers (3 plus a parent pointer), 
in addition to the space for a symbol.  Other approaches  
include the \emph{double-array trie (DAT)}~\cite{DBLP:journals/tse/Aoe89a}, which uses a minimum of 
two integers per node, each of magnitude $O(n)$.
Since a pointer must asymptotically 
use $\Omega(\log n)$ bits of memory, the asymptotic space bound
of TST (or DAT) is $O(n (\log n + \log \sigma))$ bits\footnote{Logarithms are to base 2 unless stated otherwise.}.
However, 
the information-theoretic space lower bound of $n \log \sigma + O(n)$ 
bits (see e.g. \cite{DBLP:journals/algorithmica/BenoitDMRRR05}) 
corresponds to one symbol and $O(1)$ \emph{bits} per node, so
the space usage of both TST and DAT is
asymptotically non-optimal. In practice,
$\log \sigma$ is a few bits, or one or two bytes at most. 
An overhead of $4$ pointers per node,
or $32n$ bytes on today's machines, makes it impossible 
to hold tries with even moderately many nodes in main memory.  Although
tries can be \emph{path-compressed} by deleting nodes with just
one child and storing paths explicitly, this approach (or more elaborate ones \cite{DBLP:journals/algorithmica/NilssonT02}) 
cannot guarantee a small space bound.

Motivated by this, a number of space-efficient solutions
were proposed \cite{DBLP:conf/focs/Jacobson89,DBLP:journals/algorithmica/BenoitDMRRR05,DBLP:journals/talg/RamanRS07,DBLP:journals/algorithmica/FarzanM14,DBLP:conf/icalp/FarzanRR09}, which represent
\emph{static} tries in information-theoretically optimal space,
and support a wide range of operations.  
A number of asymptotic worst-case results for dynamic tries are given in 
\cite{DBLP:conf/soda/MunroRS01,DBLP:conf/icalp/RamanR03,raey,DBLP:journals/algorithmica/JanssonSS15}.  As our focus is on practical performance,
we do not discuss all previous results in detail and refer the
reader to, e.g., \cite{raey}, for a comparison.  For completeness,
we give a summary of some of the results of 
\cite{raey,DBLP:journals/algorithmica/JanssonSS15}.  These results, 
use the standard word RAM model with word size $\Theta(\log n)$ bits,
as do ours. The first uses almost optimal $2n + n \log \sigma + o(n \log \sigma)$ bits, and supports trie operations in $O(1)$ time if 
$\sigma = {\mathrm{polylog}}(n)$ and in $O(\log \sigma/\log \log \sigma)$
time otherwise.  The second~\cite{DBLP:journals/algorithmica/JanssonSS15}
uses $O(n \log \sigma)$ bits and supports individual dynamic
trie operations in $O(\log \log n)$ amortized expected time, 
although finding the longest prefix of a string in the trie can be 
done in $O(\log k/\log_\sigma n + \log \log n)$ expected time. 
Neither of these methods has been fully implemented, although a
preliminary attempt (without memory usage measurements) 
was presented in~\cite{WAAC13}. Finally, we mention the \emph{wavelet trie}~\cite{DBLP:dblp_conf/pods/GrossiO12}, 
which is a data structure for a sequence of strings, and in principle
can replace tries in many applications. Although in theory it is dynamic,
we are not aware of any implementation 
of a dynamic wavelet trie.

Predating most of this work,
Darragh et al.~\cite{DBLP:journals/spe/DarraghCW93} 
proposed the \emph{Bonsai} data structure,
which uses a different approach to support the above dynamic trie
operations in $O(1)$ expected time (based on assumptions about
the behaviour of hash functions, which we will describe later).
The Bonsai data structure, although displaying excellent practical performance in terms of run-time, has
some deficiencies as a dynamic trie data structure.  
It is expressed in terms of a parameter $M$, which we will
refer to as the \emph{capacity} of the trie.   In what follows, unless
stated explicitly otherwise, $M$ will be used to refer to this parameter, 
$n$ (which must satisfy $n < M$) will refer to the current 
number of nodes in the trie, $\epsilon$ will be used to denote $(M-n)/M$ 
(i.e. $\epsilon$ satisfies $n = (1-\epsilon)M$), and $\alpha$ will be
used to denote $1-\epsilon$.

The Bonsai data structure allows 
a trie to be grown from empty to 
a theoretical maximum of $M$ nodes, uses 
$M \log \sigma + O(M \log \log M)$ bits of space
and performs the above set of operations in $O(1)$ expected time.
However:
\begin{itemize}
\item It is clear that, to perform operations in $O(1)$ expected time and
use space that is not too far from optimal, $n$ must lie between
$(1-c_1) M$ and $(1-c_2) M$, for small constants
$0 < c_2 < c_1 < 1$.  The standard way to 
maintain the above invariant is by periodically 
rebuilding the trie with a new smaller or larger value of $M$, 
depending on whether $n$ is smaller than $(1-c_1) M$ or 
larger than $(1-c_2) M$, respectively.  

To keep the expected (amortized) cost of rebuilding down to $O(1)$, 
the rebuilding must take $O(n)$ expected time.  
We are not aware of any way to achieve this without
using $\Theta(n \log n)$ additional bits---an unacceptably 
high cost.  The natural approach to rebuilding,
to traverse the old tree and copy it node-by-node to the new data structure,
requires the old tree to be traversed in $O(n)$ time.  Unfortunately,
in a Bonsai tree only root-to-leaf 
and leaf-to-root paths can be traversed efficiently.
While a Bonsai tree can be traversed in $O(n \sigma)$ time, this is
too slow if $\sigma$ is large.

\item Even if the above relationship between $n$ and $M$ is maintained, 
the space is non-optimal due to the additive $O(M \log \log M)$ bits term
in the space usage of Bonsai, which can dominate the remaining
terms of $M (\log \sigma + O(1))$ bits when $\sigma$ is small.
\end{itemize}
In addition, the Bonsai data structure also has a certain chance of failure:
if it fails then the data structure may need to be rebuilt, and
its not clear how to do this without affecting the space and
time complexities. Finally, it is not clear how to support
$\delLeaf$ (Darragh et al. do not claim to support this operation).

In this paper, we introduce m-Bonsai\footnote{This could be read as mame-bonsai (\begin{CJK}{UTF8}{min}豆盆栽\end{CJK}) a kind of small bonsai plant, or mini-bonsai.}, a variant of Bonsai. The advantages of m-Bonsai include:

\begin{itemize}
\item Based upon the same assumptions about the behaviour of 
hash functions as in Darragh et al. \cite{DBLP:journals/spe/DarraghCW93}, 
our variant uses $M (\log \sigma + O(1))$ bits of memory in expectation, 
and supports $\getChild$ in $O(1/\epsilon)$ expected
time, and $\getParent$, $\getRoot$, and $\getLabel$ in $O(1)$ time.

\item Using $O(M \log \sigma)$ additional bits of space, we are able to
traverse a Bonsai tree in $O(n + M)$ time (in fact, we can traverse
it in sorted order within these bounds).

\item $\addLeaf$ and $\delLeaf$ can be supported
  in $O((1/\epsilon)^2)$ amortized
  expected time.  Note that in
$\delLeaf$ the application needs to ensure that a deleted node is indeed
a leaf\footnote{If the application cannot ensure this, one can use %
the \emph{CDRW array} data structure~\cite{JSS12,DBLP:conf/alenex/PoyiasPR17} %
to maintain the number of children 
of each node (as it changes with insertions and deletions)  %
with no asymptotic slowdown in time complexity and an additional %
space cost of $O(n \log \log n)$ bits.}.
\end{itemize}
As a consequence, we obtain a trie representation that, for any
given constant $\beta > 0$, uses at most 
$(1 + \beta) (n \log \sigma + O(n))$ bits of space, and supports
the operations $\addLeaf$ and $\delLeaf$ in $O((1/\beta)^2)$
expected amortized time, $\getChild$ in $O(1/\beta)$ expected time,
and $\getLabel$ and $\getParent$ 
in $O(1)$ worst-case time. This trie representation,
however, periodically uses $O(n \log \sigma)$ bits of 
temporary additional working memory to traverse and rebuild the trie.

We give two practical variants of m-Bonsai.
Our implementations and experimental evaluations show 
our first variant to be consistently faster,
and significantly more space-efficient,
than the original Bonsai.  The 
second variant is even more space-efficient but rather slower.
Of course, all Bonsai variants use at least
20 times less space than TSTs for small alphabets and compare well 
in terms of speed with TSTs.  We also note that our experiments
show that the hash functions used in Bonsai appear to behave in
line with the assumptions about their behaviour.

The rest of this paper is organized as follows.
In Section 2, we summarize the Bonsai data structure~\cite{DBLP:journals/spe/DarraghCW93}, focussing on asymptotics.
Section 3 summarizes the m-Bonsai approach.
In Section 4 we give practical approaches to
implementing the data structure described in Section 3, 
and give details of the implementation and experimental evaluation.

%% file: preliminaries.tex
\section{Preliminaries}
\label{sec:prelim}



\subsection{Bit-vectors.}
Given a bit string $x_1,\ldots,x_n$, we define the following operations:
\begin{description}
\item[$\selectOne(x,i)$:] Given an index \textit{i}, return the location of $i_{th}$ \texttt{1} in \textit{x}. 
\item[$\rankOne(x,i)$:] Return the number of 
\texttt{1}s upto and including location \textit{i} in \textit{x}.
\end{description}

\begin{lemma}[P\v{a}tra\c{s}cu \cite{DBLP:conf/focs/Patrascu08}]
\label{lem:rankselect}
A bit string can be represented in $n + O(n/(\log n)^2)$ bits such that
$\selectOne$ and $\rankOne$ can be supported in $O(1)$ time.
\end{lemma}
The above data structure can be dynamized by adding the following operation:
\begin{description}
\item[$\flip(x,i)$:] Given an index \textit{i}, set $x_i$ to $1-x_i$.
\end{description}
It can be shown that for bit-strings that are polynomial in the word size, all
the above operations can be supported in $O(1)$ time:
\begin{lemma}[\cite{RRR01}]
\label{lem:rankselectdynamic}
A bit string of length $m = w^{O(1)}$ bits can be represented in $m + o(m)$ bits such that
$\selectOne$, $\rankOne$ and $\flip$ can be supported in $O(1)$ time, where $w$ is
the word size of the machine.
\end{lemma}

%
%
%

\subsection{Compact hash tables}
We now sketch the \emph{compact hash table (CHT)} described by
Cleary \cite{DBLP:journals/tc/Cleary84}.  The objective is
to store a set $X$ of key-value pairs, where the keys are from
$U = \{0,\ldots, u-1\}$ and the values (also referred to
as \emph{satellite data}) are from $\Sigma = \{0,\ldots,\sigma-1\}$,
and to support insertion and deletion of key-value pairs, checking membership
of a given key in $X$, and retrieval of the value associated with a given key.

Assuming that $|X| = n$, we let $M > n$ be the \emph{capacity} of the CHT.
The CHT is comprised
of two arrays, $Q$ and $F$ and two bit-strings of length $M$.
To be notationally consistent with the introduction, we take $\epsilon = (M-n)/M$.
The description below assumes that $n$ does not change
significantly due to insertions and deletions into $X$---standard 
techniques such as rebuilding can be used to avoid this assumption. 
Keys are stored in $Q$ using open addressing and linear probing\footnote{A variant, \emph{bidirectional} probing, is used in \protect{\cite{DBLP:journals/tc/Cleary84}}, but we simplify this to linear probing.}.
In normal implementations of hashing with open addressing, the keys would be stored
in $Q$ so that they can be compared with a key that is the argument to a given operation.
$Q$ would then take $M \lceil \log U \rceil$ bits, which is excessive.
To reduce the space usage of $Q$, Cleary uses
\emph{quotienting} \cite{Knuth:1998:ACP:280635}.
In rough terms, the aim of quotienting is the following:
given a hash function $h : U \rightarrow M$, a good hash function $h$
will have $O(u/M)$ elements of $U$ that map to any given $i \in M$.
Let $x \in U$ be a key that already exists in the hash table,
such that $h(x) = i$. For $x$, it should suffice to use a \emph{quotient} value $q(x)$ of
$\log(u/M) + O(1)$ bits to distinguish $x$ from any other keys that may map to $i$.  To represent
$x$ in in the CHT, we store $q(x)$ in $Q$, and
during any operation, given $q(x)$ and $h(x)$, we should
be able to reconstruct $x$ in $O(1)$ time.  If a hash function $h$ supports this
functionality, we say that it \emph{supports quotienting}.

While checking for membership of $x$ in $X$, one needs to know $h(y)$ for all
keys $y$ encountered during the search. 
As keys may not be stored at their
initial addresses due to collisions,  
the idea is to keep 
all keys with the same initial address in consecutive
locations in $Q$ (this means that keys may be moved after they have been
inserted) and to use the two bit-strings to effect
the mapping from a key's initial address to the position in $Q$ containing
its quotient (for details of the bit-strings see \cite{DBLP:journals/tc/Cleary84}). 
The satellite data is stored in an array $F$ that has 
entries of size $\lceil \log \sigma \rceil$. 
The entry $F[i]$ stores the data associated with the key whose
quotient is stored in $Q[i]$ (if any).  
 Thus, the overall space usage is $M (\log (u/n) + \log \sigma + O(1))$, which
is within a small constant factor of the information-theoretic
minimum space of $\log {{u}\choose{n}} + n \log \sigma = n \log(u/n) + n \log\sigma n \log e - \Theta(n^2/u)$ bits. To summarize:

\begin{theorem}[\cite{DBLP:journals/tc/Cleary84}]
\label{thm:compacthash}
There is a data structure that stores a set $X$ of $n$ keys from
$\{0,\ldots,u-1\}$ and associated satellite data from 
$\{0,\ldots,\sigma-1\}$ in at most
$M (\log (u/n) + \log \sigma + O(1))$ bits, for any value $M > n$
and supports insertion, deletion, membership
testing of keys in $X$ and retrieval of associated satellite data in
$O(1/\epsilon)$ time, where $\epsilon = (M - n)/M$. This assumes 
the existence of a hash function that supports quotienting and
satisfies the full randomness assumption.
\end{theorem}

\begin{remark}
  The full randomness assumption \cite{DBLP:conf/stacs/Dietzfelbinger12} states that
  for every $x \in U$, $h(x)$ is a random variable that is (essentially) uniformly
  distributed over $\{0,\ldots,M-1\}$ and is independent of the random variable
  $h(y)$ for all $y \not = x$.

  We now give an example of a hash function that supports quotienting that was used
by Cleary. The hash function has the form  $h(x) = {{({ax} \bmod {p})} \bmod {M}}$ for some prime 
$p > u$ and multiplier $a$, $1 \le a \le p-1$.  
For this $h$ we set $q(x) = \lfloor ({ax} \bmod {p}) / M \rfloor$ 
corresponding to $x$. Given $h(x)$ and $q(x)$, it is possible to
reconstruct $x$ to check for membership.  Thus, $h$ supports quotienting.
Observe that $q(x)$ is at most $\lceil \log \lfloor (p-1) / M \rfloor$.  In addition,
hash the table needs to store one additional value to indicate an empty (unused) location.
Thus, the values in $Q$ are at most $\lceil \log \lfloor (p-1) / M \rfloor + 2 \rceil$ bits.
Since we can always find a $p \le 2u$, the values in $Q$ are at most $\log(u/n) + 3$ bits. Although this family of hash functions is known not to satisfy the full randomness assumption, our implementation uses this family as well, and it performs well in practice in the applications that we consider.
\end{remark}

\subsection{Asymptotics of Bonsai}
We now sketch the Bonsai data structure,
focussing on asymptotics.  Let $M$ be the
capacity of the Bonsai data structure, $n$ the current number
of nodes in the Bonsai data structure, and let $\epsilon = (M-n)/M$ 
and $\alpha = (1-\epsilon)$ be as defined in the Introduction.  
The Bonsai data structure uses a CHT with capacity $M$,
and refers to nodes via a unique \emph{node ID}, which is
a pair $\langle i, j \rangle$ where $0 \le i < M$ and
$0 \le j < \lambda$, where $\lambda$ is an integer parameter
that we discuss in greater detail below.  If we wish to add a
child $w$ with symbol $c \in \Sigma$ to a node $v$ with node ID 
$\langle i, j\rangle$, then $w$'s node ID is obtained as 
follows: We create the \emph{key} of $w$ using
the node ID of $v$, which is a triple $\langle i, j, c\rangle$.
We insert the key into the CHT.  Letting 
$h: \{0, \ldots, M \cdot \lambda \cdot \sigma -1 \} \mapsto \{0,\ldots,M-1\}$
be the hash function used in the CHT.  We evaluate $h$
on the key of $w$. If 
$i' = h(\langle i, j, c\rangle)$,
the node ID of $w$ is $\langle i', j' \rangle$ where $j' \ge 0$ is the lowest
integer such that there is no existing node with a node ID 
$\langle i', j' \rangle$; $i'$ is called the \emph{initial address} of $w$.

Given the ID of a node, we can search for the key 
of any potential child in the CHT, which allows us to check 
if that child is present
or not.  However, to get the node ID of the child, we need to recall that all
keys with the same initial hash address in the CHT are stored consecutively.  The
node ID of the child is obtained by checking its position within the set of keys
with the same initial address as itself.  This explains how to support the operations
$\getChild$ and $\addLeaf$; for $\getParent(v)$ note that the key of $v$ encodes the
node ID of its parent. There are, however, two weaknesses in the functionality of
the Bonsai tree.

\begin{itemize}
\item It is not clear how to support
$\delLeaf$ in this data structure without affecting the time and space complexities
(indeed, as stated earlier, Darragh et al. \cite{DBLP:journals/spe/DarraghCW93} 
do not claim support for $\delLeaf$).  If a leaf 
$v$ with node ID $\langle i, j \rangle$ is deleted, we may not be able to 
handle this deletion explicitly by moving keys to close the gap, as 
any other keys with the same initial address \emph{cannot be moved} without changing
their node IDs (and hence the node IDs of \emph{all} their descendants).  Leaving a gap by
marking the location previously occupied by $v$ as
``deleted'' has the problem that the newly-vacated location 
can only store keys that have the same
initial address $i$ (in contrast to normal open address hashing).  
To the best of our knowledge, there is no analysis of 
the space usage of open hashing under this constraint.

\item
As noted in the Introduction, is not obvious how to traverse an $n$-node tree in
better than $O((n \sigma)/\epsilon)$ time. 
This also means that the 
Bonsai tree cannot be resized if $n$ falls well below (or comes too close to) 
$M$ without affecting the overall time complexity of $\addLeaf$ and $\delLeaf$.
\end{itemize}

\paragraph{Asymptotic space usage.}
\label{sec:AsymptoticBons}
In addition to the two bit-vectors of $M$ bits each, the main
space usage of the Bonsai structure is $Q$.  
Since a prime $p$ can be found that is 
$< 2\cdot M \cdot \lambda \cdot \sigma$, 
it follows that the values in $Q$ 
are at most $\lceil \log (2 \sigma \lambda + 1) \rceil$ bits.
The space usage of Bonsai is therefore
$M (\log \sigma + \log \lambda + O(1))$ bits. 

Since the choice of the prime $p$ depends on $\lambda$, 
$\lambda$ must be fixed in advance.  However, if more than $\lambda$ keys
are hashed to any value in $\{0,\ldots,M-1\}$, the
algorithm is unable to continue and must spend $O((n \sigma)/\epsilon)$ time 
to traverse and possibly re-hash the tree.
Thus, $\lambda$ should be chosen large enough to reduce the probability 
of more than $\lambda$ keys hashing to the same initial address
to acceptable levels.  In \cite{DBLP:journals/spe/DarraghCW93} the authors, assuming the hash function satisfies the full randomness assumption
argue that choosing $\lambda = O(\log M /\log \log M)$ 
reduces the probability of error to at most $M^{-c}$ for any constant $c$ 
(choosing asymptotically smaller $\lambda$ causes the
algorithm almost certainly to fail). 
As the optimal space usage for an $n$-node trie on an
alphabet of size $\sigma$ is $O(n \log \sigma)$ bits,
the additive term of $O(M \log \lambda) = O(n \log \log n)$ makes
the space usage of Bonsai non-optimal for small alphabets.

However, even this choice of $\lambda$ is not
well-justified from a formal perspective, since
the hash function used is quite weak---it is only 2-universal 
\cite{DBLP:journals/jcss/CarterW79}.
For 2-universal hash functions, the maximum number of collisions 
can only be bounded to $O(\sqrt{n})$ \cite{DBLP:journals/jacm/FredmanKS84}
(note that it is not obvious
how to use more robust hash functions, since quotienting may not be possible).
Choosing $\lambda$ to be this large would make the space usage of the
Bonsai structure asymptotically uninteresting.

\paragraph{Practical analysis.}
In practice, we note that choosing $\lambda = 16$ as 
suggested in \cite{DBLP:journals/spe/DarraghCW93} gives a relatively 
high failure probability for $M = 2^{56}$ and $\alpha = 0.8$, 
using the formula in \cite{DBLP:journals/spe/DarraghCW93}.   
Choosing choosing $\lambda = 32$, and assuming the hash function
satisfies full randomness, the error probability
for $M$ up to $2^{64}$ is about $10^{-19}$ for $\alpha = 0.8$
Also, in practice, the prime $p$ is not significantly larger than 
$M \lambda \sigma$ \cite[Lemma 5.1]{DBLP:journals/siamcomp/Pagh01}.
As a result (see the paragraph following \ref{thm:compacthash}) the space usage of Bonsai is typically well approximated by $M (\lceil \log \sigma \rceil + 7)$
bits for the tree sizes under consideration in this paper (for
alphabet sizes that are powers of 2, we should replace the 7 by 8).

%% file: bonsai_new.tex
\subsection{Collision handling using $O(M)$ bits}
In our approach, 
each node again has an associated key
that needs to be searched for in a hash table, again
implemented using open addressing with linear probing and
quotienting.  However, the ID of a node $x$ in our case is 
a number from $\{0,\ldots, M-1\}$ that refers to the index 
in $Q$ that contains the quotient corresponding to $x$. If a
node with ID $i$ has a child with symbol $c \in \Sigma$,
the child's key, which is $\langle i, c \rangle$, is hashed
using a multiplicative hash function 
$h: \{0,\ldots,M\cdot \sigma - 1\} \mapsto \{0,\ldots, M-1\}$, and an
initial address $i'$ is computed.  If $i''$ is the
smallest index $\ge i'$ such that $Q[i'']$ is vacant, then we
store $q(x)$ in $Q[i'']$.  Observe that $q(x) \le \lceil 2 \sigma \rceil$,
so $Q$ takes $M \log \sigma + O(M)$ bits. In addition, we have
a \emph{displacement} array $D$, and set $D[i''] = i'' - i'$.
From the pair $Q[l]$ and $D[l]$, we can obtain both the
initial hash address of the key stored there and its quotient,
and thus reconstruct the key.
The key idea is that in expectation, the average value
in $D$ is small:

\begin{proposition}
\label{prop:space}
Assuming $h$ is fully independent and uniformly random,
the expected value of $\sum_{i=0}^{M-1} D[i]$ after all $n = \alpha M$ nodes have
been inserted is $\approx M \cdot \frac{\alpha^2}{2(1-\alpha)}$.
\end{proposition}
\begin{proof}
The average number of probes, 
over all keys in the table, made in a successful search is
$\approx \frac{1}{2}(1 + \frac{1}{1-\alpha})$ \cite{Knuth:1998:ACP:280635}. Multiplying
this by $n = \alpha M$ gives the total average number of probes.
However, the number of probes for a key is one more than its
displacement value.  Subtracting $\alpha M$ from the above and
simplifying gives the result.
\end{proof}
Thus, encoding $D$ using variable-length encoding
could be very beneficial.  For example, coding $D$ in unary would
take $M + \sum_{i=1}^M D[i]$ bits; by Proposition~\ref{prop:space}, and plugging
in $\alpha = 0.8$, the expected space usage of $D$, encoded in unary, 
should be about $2.6M$ bits, which is smaller
than the overhead of $7M$ bits of the original Bonsai.
As shown in Table~\ref{tab:GammaGolombUnary}, predictions made
using Proposition~\ref{prop:space} are generally quite accurate. 
Table~\ref{tab:GammaGolombUnary} also suggests that 
encoding each $D[i]$ using the Elias-$\gamma$-code~\cite{elias:ref} 
(hereafter abbreviated to just $\gamma$-code), 
we would come down to about $2.1M$ bits for the $D$, for $\alpha = 0.8$.

\setlength{\intextsep}{0.5pt plus 1.0pt minus 2.0pt}
\begin{table}
\caption{Average number of bits per entry needed to encode the displacement array using the unary, Elias-$\gamma$ and Golomb encodings. For the unary encoding, Proposition~\ref{prop:space} predicts $1.81\dot{6}$, $2.6$ and $5.05$ bits per value. For file details see Table 2.}
\label{tab:GammaGolombUnary}

\centering
\begin{tabular}{|r||r|r|r||r|r|r||r|r|r||}
\hline
 & \multicolumn{3}{|c||}{unary} & \multicolumn{3}{|c||}{$\gamma$} & \multicolumn{3}{|c||}{Golomb}\\ \hline
 Load Factor & 0.7 & 0.8 & 0.9 & 0.7 & 0.8 & 0.9 & 0.7 & 0.8 & 0.9 \\ \hline
 Pumsb & 1.81 & 2.58 & 5.05  & 1.74 & 2.11 & 2.65 & 2.32 & 2.69 & 3.64 \\ 
 Accidents & 1.81 & 2.58 & 5.06 & 1.74 & 2.11 & 2.69 & 2.33 & 2.69 & 3.91 \\ 
 Webdocs & 1.82 & 2.61 & 5.05 & 1.75 & 2.11 & 2.70 & 2.33 & 2.70 & 3.92 \\ \hline
\end{tabular}
\end{table}
\setlength{\intextsep}{0.5pt plus 1.0pt minus 2.0pt}

\subsection{Representing the displacement array}

We now describe how to represent the displacement array.
A \emph{compact dynamic rewriteable array (CDRW array)} 
is a data structure for a sequence of
non-negative integers that supports the following operations:

\begin{description}
\item[$\create(n)$:] Create an array $A$ of size $n$ with all entries initialized to zero.
\item[$\set(A, i, v)$:] Set $A[i]$ to $v$ where $v \ge 0$ and $v = n^{O(1)}$.
\item[$\get(A, i)$:] Return $A[i]$.
\end{description}
The following lemma shows how to implement such a data structure.  
Note that the apparently slow running time of $\set$ is
enough to represent the displacement array without asymptotic
slowdown: setting $D[i] = v$ means that $O(v)$ time has already
been spent in the hash table finding an empty slot for the key.

\begin{lemma}
\label{lem:woda}
A CDRW array $A$ of size $n$ 
can be represented in
$\sum_{i=1}^n |\gamma(A[i] + 1)| + n + o(n)$ bits, supporting 
$\get$ in $O(1)$ time  and $\set(A, i, v)$ in $O(v)$ amortized time.
\end{lemma}

\begin{proof}
We divide $A$ into contiguous blocks of size $b = (\log n)^{3/2}$.
The $i$-th block $B_i = A[bi..bi+b-1]$ will be stored in one or
two contiguous sequences of memory locations. There will be a pointer 
pointing to the start each of these contiguous sequences of
memory locations.

We first give a naive representation of a block.
All values in a block are encoded using $\gamma$-codes and
(in essence) concatenated into a single bit-string.
A $\set$ operation is performed
by decoding all the $\gamma$-codes in the block, and re-encoding
the new sequence of $\gamma$-codes.  Since each 
$\gamma$-code is $O(\log n)$ bits, or $O(1)$ words, long, it
can be decoded in $O(1)$ time.  Decoding and re-encoding
an entire block therefore takes $O(b)$ time, which is also
the time for the $\set$ operation.
A $\get$ operation can be realized in $O(1)$ time using
the standard idea of to concatenating the unary and binary portions
of the $\gamma$-codes separately into two bit-strings, and to 
use $\selectOne$ operations on the unary bit-string to obtain, 
in $O(1)$ time, the binary portion of the $i$-th $\gamma$-code.
Let $G_i = \sum_{j=bi}^{bi+b-1} |\gamma(A[j]+1)|$.
The space usage of a single block is therefore
$G_i + O(G_i/(\log n)^2 + \log n)$ bits, where 
the second term comes from Lemma~\ref{lem:rankselect} applied to
the unary part of the $\gamma$-codes and the third 
accounts for the pointers to the block and any unused space in
the ``last'' word of a block representation.
The overall space usage of the naive representation is therefore
$\sum_i G_i + O((\sum_i G_i)/(\log n)^2) + (n \log n)/b)$ bits: 
this adds up to $\sum_i G_i + o(n)$ bits as required.  The problem,
however, is that $\set$ takes $O(b) = O((\log n)^{3/2})$ time,
regardless of the value being set.

To improve the speed of the $\set$ operation, we first
make the observation that by a standard amortization argument,
we can assume that the amortized time allowed for $\set(A, i, v)$ is $O(v+v')$,
where $v'$ is the current value of $A[i]$.  Next, we classify the values
in a block as being either large or small, depending on whether
or not they are $\ge b$. Each block is now stored as two bit-strings,
one each for small and large values.  An additional bit-string,
of length exactly $b$ bits, indicates whether a value in a block
is large or small.  By performing $\rankOne$ and $\flip$
operations in $O(1)$ time on this bit-string using Lemma~\ref{lem:rankselectdynamic}
we can reduce a $\set$ or $\get$ operation on the block to a
$\set$ or $\get$ operation on either the large or the small bit-strings.

The bit-string for large
values (the large bit-string) 
is represented as in the naive representation. The space usage
for the large bit-string for the $i$-th block is 
$G^{l}_i + O(G^l_i/(\log n)^2 + \log n)$ bits, 
where 
 $G^l_i = \sum_{\{bi \le j < bi+b|A[j] \mbox{\footnotesize{\ is large}}\}} |\gamma(A[j]+1)|$. Observe
that if either $v'$ or $v$ is large, then the $O(b)$ time needed to
update the large bit-string is in fact $O(v+v')$ as required.

We now consider the representation of the small bit-string.
Note that all $\gamma$-codes in the small bit-string 
are $O(\log b) = O(\log \log n)$ bits long.
We divide a block into \emph{segments} of 
$\lambda = \lceil c \log n/\log \log n \rceil$ values
for some sufficiently small constant $c>0$.
All small values in a segment are stored as a concatenation
their $\gamma$-codes, followed by an 
\emph{overflow} zone of between $o/2$ and $2o$ bits,
where $o = \lceil \sqrt{\log n}\log \log n \rceil$ 
bits (initially, all overflow zones are of size $o$ bits). 
All segments, and their overflow zones, are concatenated
into a single bit-string. The small bit-string of the $i$-th block,
also denoted $B_i$, has length at most 
$G^s_i + (b/\lambda) \cdot (2o) = G^s_i + O(\log n (\log \log n)^2)$,  
where  $G^s_i = \sum_{\{bi \le j < bi+b|A[j] \mbox{\footnotesize{\ is small}}\}} |\gamma(A[j]+1)|$. We now discuss how to support operations on $B_i$.

\begin{itemize}
\item Firstly, we need to be able to find the starting and ending points
of individual segments in a block.  As the size of a segment 
and its overflow zone is an integer of at most $O(\log \log n)$ bits,
and there are only $O(\sqrt{\log n}\log \log n)$ segments in a block,
we can store the sizes of the segments and overflow blocks in a block 
in a single word and perform the appropriate prefix sum operations 
in $O(1)$ time using table lookup.

\item As we can ensure that a segment is of size at most $(\log n)/2$
by choosing $c$ small enough, we can support a $\set$ operation on 
a segment in $O(1)$ time, by overwriting the sub-string of $B_i$ that 
represents this segment (note that if $v$, new value, is large, then
we would need to excise the $\gamma$-code from the small bit-string,
i.e. replace its old $\gamma$-code by an empty string).
If the overflow zone exceeds $2o$ or goes below $o/2$ bits, 
the number of $\set$ operations that have taken place in this segment alone
is $\Omega(\sqrt{\log n}\log \log n)$.  
Since the length of $B_i$ is at most $O(b \log \log n)$ bits or 
$O(\sqrt{\log n} \log \log n)$ words, when any segment overflows, 
we can simply copy $B_i$ to a new sequence of memory words, 
and while copying, use table lookup
again to rewrite $B_i$, ensuring that each segment has 
an overflow zone of exactly $o$ bits following it.
Note that an overflow zone can only change in size by a constant
factor, so rewriting a section of $c \log n$ bits in $B_i$ will give
a new bit-string which is also $O(\log n)$ bits long, and the rewriting
takes $O(1)$ time as a result.  
The amortized cost of rewriting is clearly $O(1)$, so the
case where a large value is replaced by a small one, thus necessitating
a change in $B_i$ even though the update is to the large bit-string
is also covered.

\item To support a $\get$ in $O(1)$ time is straightforward.  
After finding the start of a segment, we extract the appropriate
segment in $O(1)$ time and read and decode the appropriate $\gamma$-code
using table lookup.
\end{itemize}
Adding up the space usages for the large and small blocks and the bit-string
that distinguishes between large and small values gives Lemma~\ref{lem:woda}.
\end{proof}

\subsection{An alternative representation of the displacement array}
\label{subsec:darray-alt}

A quick inspection of the proof of Lemma~\ref{lem:woda} shows that the data
structure is not especially practical.  We now give an alternative 
approach that is,  asymptotically speaking, slightly less space-efficient.
In this approach, we choose two integer parameters
$1 < \Delta_0 < \Delta_1$, and 
the displacement array $D$ is split into three layers: 

\begin{itemize}
\item The first layer consists of an array $D_0$ of \emph{equal-length} entries 
of $\Delta_0$ bits each, which has size $M_0 = M$.  
All displacement values $\le 2^{\Delta_0}-2$ are stored as is in $D_0$.  
If $D[i] > 2^{\Delta_0}-2$, then we set $D_0[i] = 2^{\Delta_0} -1$.

\item The second layer consists of a CHT with maximum size $M_0 \le M$.
If ${2^{\Delta_0}-1 \leq D_1[i] \leq 2^{\Delta_0} +2^{\Delta_1}-2}$, then we store the 
value $D[i] - 2^{\Delta_0} + 1$ as satellite data associated with the 
key $i$ in the second layer. Note that the satellite data has
value between $0$ and $2^{\Delta_1} -1$ and so fits into $\Delta_1$ bits.

\item The third layer consists of a standard hash table.
If $D[i] > 2^{\Delta_0} +2^{\Delta_1}-2$, we store $D[i]$ in this
hash table as satellite date associated with the key $i$.
\end{itemize}

Clearly, $D[i]$ can be accessed in $O(1)$ expected time.
We now describe how to choose the parameters
$\Delta_0$ and $\Delta_1$.  We use the following theorem,
which is an adaptation of a textbook proof of the expected search
cost of linear probing \cite{GoodrichTamassiaBook}:
\begin{theorem}
\label{thm:CLRS}
  Given an open-address hash table with load factor $\alpha = n/M < 1$,
  assuming full randomness, 
  the probability that an unsuccessful search makes $\ge k$ probes
  is at most $b_\alpha \cdot c_{\alpha}^k$ for some
  constants $b_{\alpha}$ and $c_\alpha < 1$ that depend only on $\alpha$.
\end{theorem}
\begin{proof}
  Let $A[0..M-1]$ be the hash table. Let $x$ be a key
  that is not in the hash table, and let $h(x) = i$.
  If the search for $x$ makes $\ge k$ probes then the locations
  $A[i],A[i+1],\ldots,A[i+k-1]$ must be occupied.\footnote{To simplify notation we ignore wrapping around the ends of $A$.}
  A necessary condition for this to happen is that there must exist
  a $k' \ge k$ such that $k'$ keys are mapped to
  $A[i-k'+k], \ldots, A[i+k-1]$, and the remaining
  keys are mapped to $A[0..i-k'+k-1]$ or $A[i+k..M-1]$.
  Under the assumption of full randomness,
  the number of keys mapped to $A[i-k'+k], \ldots, A[i+k-1]$
  is binomially distributed with parameters $k'/M$ and $n$, and
  the expected number of keys mapped to this region is
  $k' n/M = \alpha k'$. Using the multiplicative form of the
  Chernoff bound, we get that the probability of $\ge k'$ keys
  being hashed to this region is at most:
  $$\left ( \frac{e^{1 - \alpha}}{(1/\alpha)^{1/\alpha}}  \right )^{k'} = c^{k'}_\alpha,$$
    where $c_{\alpha} < 1$ is a constant that depends only on $\alpha$.
    Summing over $k' = k, k+1,\ldots, n$ we get that the probability
    of an unsuccessful search taking over $k$ probes is
    at most $b_\alpha \cdot c^k_\alpha$, as desired.
  %
\end{proof}

We now analyze the space usage:
\begin{itemize}
\item 
The space usage of the first layer is $M \Delta_0$. We
will choose  $\Delta_{0}=O(\log^{(5)} n)$ so the
space usage of this layer is $O(n \log^{(5)} n)$ bits.

\item Let the expected number of displacement values stored in the 
  second layer be $n_1$.  Only displacement
  values $\ge \theta_1 = {2^{\Delta_{0}}-1}$ will be stored in the second
  layer, so by Theorem~\ref{thm:CLRS}, $n_1\leq \alpha^{\theta_1} n$.
  We choose $\Delta_0$ so that $n_1 = O(n/\log^{(3)} n)$,
  so $\theta_1 = \frac{\log^{(4)} n}{\log_{1/c_\alpha} 2}$.
  It follows that $\Delta_{0}=O(\log^{(5)} n)$, as promised above.

  We now discuss the asymptotic 
  space usage of the CHT. By Theorem~\ref{thm:compacthash}, 
  the space usage of this CHT is ${M_1(\log(M/M_1)+ \Delta_1 + O(1))}$ bits, 
  where $M=n/\alpha$ is the universe and $M_1=n_1/\alpha$ is the size
  of the CHT\footnote{For simplicity, we choose the same load factor for m-Bonsai and the CHT.}.
  Since $n_1 = O(n/\log^{(3)}  n)$, we see that  $M/M_1=O(\log^{(3)} n)$,
  and hence the asymptotic space usage of the CHT is
  ${O(\dfrac{n}{\log^{(3)} n}(\log^{(4)} n + \Delta_1 + O(1)))}$ bits. 
  We will choose $\Delta_1= O(\log^{(3)} n)$, so the space usage
  of the CHT is $O(n)$ bits.

\item  Finally, we discuss the asymptotic space usage of the third layer.
Let $n_2$ be the expected number of keys stored in the third layer.
By Theorem~\ref{thm:CLRS}, $n_2\leq \alpha^{\theta_2} n$, 
where $\theta_2 ={2^{\Delta_{1}}-3}$. 
We choose $\Delta_1$ so that
$n_2 = O(n/\log n)$, so $\theta_2 = \frac{\log^{(2)} n}{\log_{1/c_\alpha} 2}$.
It follows that $\Delta_{1}=O(\log^{(3)} n)$, as promised above.
Since the expected
space usage of the third layer is $O(n_2 \log n)$ bits, this is
also $O(n)$ bits as required.
\end{itemize}

%% file: bonsai_traverse.tex
\subsection{Traversing the Bonsai tree}
In this section, we discuss how to traverse a Bonsai tree with
$n$ nodes stored in a hash table array of size $M$ in $O(M + n)$ expected
time. We first give an approach that uses $O(M \log \sigma)$ additional bits.
This is then refined in the next section to an approach that
uses only $n \log \sigma + O(M+n)$ additional bits.  Finally,
we show how to traverse the Bonsai tree in \emph{sorted} order.

\subsubsection{Simple traversal}
\label{sec:simptrav}

Traversal involves a preprocessing step to build three simple support data structures.
The first of these is an array $A$ of $M\log\sigma$-bit integers.
The preprocessing begins by scanning array $Q$ left to right. For each non-empty entry $Q[i]$ encountered 
in the scan, we increment $A[\getParent(i)]$.
At the end of the scan, a non-zero entry $A[j]$ is the number of children of node $j$.
We then create bitvector $B$, a unary encoding of $A$, which requires $M+n + o(M+n)$ bits of space.
Observe $A[i] = \selectOne(B,i) - \selectOne(B,i-1)$ and 
$\rankZero(\selectOne(B,i))$ is the prefex sum of $A[1..i]$.
We next allocate an array $C$, of size $n$ to hold the labels for the children of each node. 
To fill $C$, we scan $Q$ a second time, and for each non-zero entry $Q[i]$ encountered, we set 
$C[\selectOne(B,\getParent(i)) - getParent(i) - A[\getParent(i)]] \leftarrow \getLabel(i)$ and decrement $A[\getParent(i)]$. 
At the end of the scan, $C[\rankZero(\selectOne(B,i))..\rankZero(\selectOne(B,i+1))]$ contains precisely the labels of the 
children of node $i$, and $A$ contains all zeroes. Note that the child labels in $C$ for a given node are not necessarily 
in lexicographical order. Preprocessing time is dominated by the scans of $Q$, which take $O(M)$ time.
Space usage is $(M+n)(\log\sigma + 1 + o(1))$ bits.

With $A$, $B$ and $C$ we are able to affect a depth first traversal of the trie, as follows.
$B$ allows us to determine the label of the first child of an arbitrary node $i$ in constant time: 
specifically, it is $C[\rankZero(\selectOne(B,i))]$. Before we visit the child of $i$ with label 
$C[\rankZero(\selectOne(B,i))]$, we increment $A[i]$, which allows us, when we return to node $i$ 
having visited its children, to determine the label of the next child of node $i$ to visit: it is 
$C[\rankZero(\selectOne(B,i)) + A[i]]$. Traversal, excluding preprocessing, clearly takes $O(n)$ time.

\subsubsection{Reducing space}

We can reduce the space used by the simple traversal algorithm described above by exploiting
the fact that the $M$ values in $A$ sum to $n$ and so can be represented in $n$ bits, in such 
a way that they can be accessed and updated in constant time with the help of $B$. Essentially,
we will reuse the $M\log\sigma$ bits initially allocated for $A$ to store the $C$ array and a 
compressed version of $A$.

We allocate $\min(M\log\sigma,M+n(1+\log\sigma))i$ bits for $A$, compute it in the manner described 
in the simple traversal algorithms, and then use it to compute $B$. The space allocated for $A$ is 
sufficient to store $C$, which is of size $n\log\sigma$ bits, and at least another $n+M$ bits. Denote 
these $n+M$ bits $A'$. We will use $B$ to divide $A'$ into variable length counters. Specifically, 
bits $A'[\selectOne(B,i-1))+1..\selectOne(B,i))]$ will be used to store a counter that ranges from 
0 to the degree of node $i$. $A'$ replaces the use of $A$ above during the traversal phase. 

\subsubsection{Sorted traversal}

We now describe a traversal that can be used to output the strings present in 
the trie in lexicographical order. In addition to the data structures used in simple traversal, we 
store $L$, a set of $\sigma$ lists of integers, one for each symbol of the alphabet.

We begin by scanning $Q$ left to right, and for each non-empty entry $Q[i]$ encountered in the scan, 
we increment $A[\getParent(i)]$ (as in the simple traversal algorithm) and append $i$ to the list for 
symbol $\getLabel(i)$. At the end of the scan the lists contain $n$ elements in total. Note also that 
the positions in the list for a given symbol are strictly increasing, and so we store them differentially 
encoded as Elias-$\gamma$ codes to reduce their overall space to $n\log\sigma$ bits. 

We then compute $B$ as in the simple traversal algorithm, and allocate space for $C$. 
Now, however, where in the simple algorithm we would make a second scan of $Q$, we scan the lists of
$L$ in lexicographical order of symbol, starting with the lexicographically smallest symbol. For each
position $j$ encountered in the scan, we access $Q[j]$ and add $getLabel(j)$ to the next available 
position in the region of $C$ containing the child labels for node $\getParent(j)$ (which we can access
as before with $B$ and $A$. The child labels in $C$ for a given node now appear in lexicgraphical order,
allowing us to affect a lexicographic traversal of the trie.

%% file: bonsai_new_conclude.tex
\subsection{Conclusion}

\begin{theorem}
For any given integer $\sigma$ and $\beta > 0$, 
there is a data structure that represents a trie on an
alphabet of size $\sigma$ with $n$ nodes, using
$(1+\beta) (n \log \sigma + O(1))$
bits of memory in expectation, and supporting $\getRoot$, $\getParent$,
and $\getLabel$ in $O(1)$ time, $\getChild$ in $O(1/\beta)$ expected
time, and $\addChild$ and $\delLeaf$ in $O((1/\beta)^2)$ amortized
expected time. 
The data structure uses $O(n \log \sigma)$ additional bits of temporary
memory in order to periodically restructure.
The expected time bounds are based upon the existence of a hash function that
supports quotienting and satisfies the full randomness assumption.
\end{theorem}
\begin{proof}
  We represent the tree using the m-Bonsai data structure, representing
  the $D$-array using Lemma~\ref{lem:woda}. Let $M$ be the current
  capacity of the hash table representing the trie.
  We choose an arbitrary location $r \in \{0,\ldots,M-1\}$ as the
  root of the tree, and set $Q[r] = 0$ (or indeed any value that
  indicates that $r$ is an occupied location) and $D[r] = 0$ as well.
  We store $r$, which is the ID of the root node, in the data structure.
  The operations are implemented
  as follows:

  \begin{description}
  \item[$\getRoot$:] We return the stored ID of the root node.
  \item[$\getChild(v,c)$:] We create the key $\langle v, c \rangle$ and search for it in the CHT.  When searching for a key, we use the quotients stored in $Q$ and $O(1)$-time access to the $D$ array to recover the keys of the nodes we encounter during linear probing.
  \item[$\addChild(v, c)$:] We create the key $\langle v, c \rangle$ and insert it into the CHT. Let $i = h(\langle<v, c\rangle)$ and suppose that $Q[j]$ is empty for some $j \ge i$.  We set $D[j] = j-i$ -- this takes $O(j-i+1)$ time, but can be subsumed by the cost of  probing locations $i,\ldots,j$.
  \item[$\getParent(v)$:] If $v$ is not the root, we use $Q[v]$ and $D[v]$ (both accessed in $O(1)$ time) to reconstruct the key $\langle v',c\rangle$ of $v$, where $v'$ is the parent of $v$ and $c$ is the label of $v$.
  \item[$\getLabel(v)$:] Works in the same way as $\getParent$.
  \item[$\delLeaf(v, c)$:] We create the key $\langle v, c \rangle$ and search f
or it in the CHT as before.  When we find the location $v'$ that is the ID of the leaf, we store a ``deleted'' value that is distinct from any quotient or from the ``unoccupied'' value (clearly, if an insertion into the CHT encounters a ``deleted'' value during linear probing, this is treated as an empty location and the key is inserted into it.
  \end{description}

  We now discuss the time complexity of the operations.
  If $n$ is the current number
  of trie nodes, we ensure that the current value of $M$
  satisfies $(1 + \beta/2) n \le M \le (1+\beta) n$ which means
  that $\epsilon = (M - n)/n$ varies between $(\beta/2) / (1 +\beta /2)$
  and $\beta / (1+ \beta)$.  Under this assumption, the value of
  $\epsilon = (M-n)/M$ is $\Theta((1+\beta)/\beta)$, and the operations
  $\addChild$, $\getChild$ and $\delLeaf$ 
  take $O(1/\epsilon) = O(1/\beta)$ (expected) time.
  If an $\addChild$ causes $n$ to go above $M/(1+\beta/2)$, we
  create a new hash table with capacity $M' = \frac{(1+3\beta/4)}{(1+\beta/2)} M$. We traverse the old tree in $O((M+n)/\beta)$ time and copy it to the
  new hash table.  However, at least $Omega(\beta n)$
  $\addChild$ operations must have occurred since last time that the
  tree was copied to its current hash table, so the amortized
  cost of copying is $O((1/\beta)^2)$.  The case
  of a $\delLeaf$ operation causing $n$ to go below $M/(1+\beta)$ is similar.
  The space complexity, by the previous discussions, is clearly
  $M (\log \sigma + O(1))$ bits.  Since $M \le (1+\beta)n$, the
  result is as claimed.
\end{proof}

%% file: experimental.tex
In this section, we first discuss details of our 
implementations.
Next we describe the implementation of a naive approach
for traversal and the simple linear-time traversal explained in
Section~\ref{sec:simptrav}.
All implementations
are in C++ and use some components from
the \texttt{sdsl-lite} library \cite{DBLP:conf/wea/GogBMP14}.
Finally we describe our machine's specifications, our datasets,
our benchmarks, and the performance of our implementations in these
benchmarks. 

\subsection{Cleary's CHT and original Bonsai}

We implemented our own version of Cleary's CHT, which 
mainly comprises three \texttt{sdsl} containers: firstly, the 
\texttt{int\_vector<>} class, which uses a fixed number of 
bits for each entry, is used for the $Q$ array.
In addition, we have two instance of the \texttt{bit\_vector} container
to represent the bit-strings used by \cite{DBLP:journals/tc/Cleary84,DBLP:journals/spe/DarraghCW93} to map a key's initial address to the position in $Q$
that contains its quotient.
The original Bonsai trie is implemented essentially
on top of this implementation of the CHT.

\subsection{Representation of the displacement array}
As noted earlier, the data structure of Lemma~\ref{lem:woda} appears to be too
complex for implementation. We therefore implemented two practical
alternatives, one based on the naive approach outlined in 
Lemma~\ref{lem:woda} and one that is based on
Section~\ref{subsec:darray-alt}.

\subsubsection{m-Bonsai ($\gamma$)} 

We now describe m-Bonsai ($\gamma$) which is is based on representing
the D-array using the naive approach of Lemma~\ref{lem:woda}.
$D$ is split into $M/b$ consecutive blocks of $b$ displacement values each.
The displacement values are stored as a concatenation of their $\gamma$-codes.
Since $\gamma$-codes are only defined for values $>1$, and there will be
some zero displacement values, we add $1$ to all displacement values.
In contrast to the description in Lemma~\ref{lem:woda}, 
to perform a $\get(i)$, we find the block containing $D[i]$, decode all the
$\gamma$-codes in the block up to position $i$,
and return the value.
To perform $\set(i, v)$, we decode the block
containing $i$, change the appropriate value and
re-encode the entire block.

In our experiments, we choose $b = 256$. 
We have an array of pointers to the blocks.
We used \texttt{sdsl}'s \texttt{encode} and
\texttt{decode} functions to encode and decode each block 
for the \set{} and \get{} operations.

\subsubsection{m-Bonsai (recursive)}
We now discuss the implementation of the alternate representation
of the displacement array discussed in Section~\ref{subsec:darray-alt},
which we call \emph{m-Bonsai (recursive)}.
$D_{0}$ is has equal-length entries of $\Delta_{0}$-bits and
is implemented as an \texttt{int\_vector<>}.
The second layer is a CHT, implemented as discussed above.
The third layer is implemented using the C++ \texttt{std::map}.

We now discuss the choice of parameters $\Delta_0$ and $\Delta_1$.
The description in Section~\ref{subsec:darray-alt}
is clearly aimed at asymptotic analysis:
the threshold $\theta_1$ above which values end up in the
second or third layers, for $n = 2^{65,536}$  and $\alpha = 0.8$,
is about $2/9$. As a result the values of $\Delta_0$ and $\Delta_1$ are
currently chosen numerically.  Specifically, we compute the
probability of a displacement value exceeding $k$ for
load factors $\alpha = 0.7, 0.8,$ and $0.9$ using the exact
analysis in \cite{Knuth:1998:ACP:280635} (however, since we are not aware of a closed-form
formula for this probability, the calculation is done numerically).
This numerical  analysis shows, for example, that for $\alpha = 0.8$,
choosing$\Delta_0 = 2$ or $3$ and $\Delta_1 = 6, 7, or 8$ give roughly the same
expected space usage.  Clearly, choosing $\Delta_0 = 3$ would give superior
performance as more displacement values would be stored in $D_0$
which is just an array, so we chose $\Delta_0 = 3$.  Given this choice,
even choosing $\Delta_1 = 7$ (which means displacement values
$\ge 134$ are stored in the third layer), the probability of
a displacement value being stored in the third layer is at most 0.000058.
Since storing a value in the \texttt{stl::map} takes 384 bits on a
64-bit machine, the
space usage of the third layer is negligible for a 64-bit machine.


\subsection{m-Bonsai traversal}

We now discuss the implementation of traversals. As discussed, the
difficulty with both Bonsai data structures is that the $\getChild$
and $\getParent$ operations only support leaf-to-root traversal.

One approach to traversing a tree with this set of operations is as follows.
Suppose that we are at a node $v$. For $i = 0, \ldots, \sigma - 1$,
we can perform $\getChild(v, i)$ to check if $v$ has a child
labelled $i$; if a child is found, we recursively traverse
this child and its descendants.  This approach takes $O(n\sigma)$ time.



The algorithm described in Section~\ref{sec:simptrav} was implemented
using \texttt{sdsl} containers as follows.
The arrays $A$ and $C$ are implemented as \texttt{sdsl} \texttt{int-vectors}
of length $M$ and $n$ respectively, with width $\lceil \log \sigma \rceil$.
The bit-vector was implemented as a \texttt{select\_support\_mcl} class
over a \texttt{bit\_vector} of length $M+n$.

\subsection{Experimental evaluation}

\subsubsection{Datasets}
We use benchmark datasets arising in frequent pattern
mining \cite{FIMI:2003:Goethals}, where each ``string'' is a
subset of a large alphabet (up to tens of thousands). 
We also used sets of short read genome strings given in the 
standard FASTQ format. These datasets have a relatively small alphabet $\sigma = 5$. Details of the datasets can be found in Table~\ref{tab:characteristics}.


\subsubsection{Experimental setup}
All the code was compiled using g++ 4.7.3 with optimization level 6.
The machine used for the experimental analysis is an Intel Pentium 64-bit machine with 8GB of main memory and a G6950 CPU clocked at 2.80GHz with 3MB L2 cache, running Ubuntu 12.04.5 LTS Linux.  To measure the resident  memory  (RES),
 \texttt{/proc/self/stat} was used. For the speed tests we measured wall clock time using \texttt{std::chrono::duration\_cast}.

\subsubsection{Tests and results}

We now give the results of our experiments,
divided into tests on the memory usage,
benchmarks for build, traverse and successful search operations.
Our m-Bonsai approaches where compared with Bonsai and Bentley's C++ TST implementation \cite{Dobbs:1998:Bentley}. The DAT implementation of \cite{DBLP:conf/coling/0001K14} was not tested since it apparently uses 32-bit integers, limiting the maximum trie size to $2^{32}$ nodes, which is not a limitation for the Bonsai or TST approaches. The tests of \cite{DBLP:conf/coling/0001K14} suggest that even with this ``shortcut'', the space usage is only a factor of 3 smaller than TST (albeit it is $\sim$ 2 times faster).

\paragraph{Memory usage.}
We compare the aforementioned data structures in terms of memory usage. 
For this experiments we set $\alpha = 0.8$ for all Bonsai data structures. 
Then, we insert the strings of each dataset in the trees and we measure the resident memory.
Table \ref{tab:MemUsage} shows the space per node (in bits).  We note that
m-Bonsai ($\gamma$) consistently uses the least memory, followed by m-Bonsai (r).
Both m-Bonsai variants used less memory than the original Bonsai.

Since the improvement of m-Bonsai over Bonsai is a reduction in space
usage from $O(n \log \sigma + n \log \log n)$ to
$O(n \log \sigma + n)$, the difference will be greatest
when $\sigma$ is small.  This can be observed in our experimental results,
where the difference in space usage between m-Bonsai and the original
Bonsai is greatest when $\sigma$ is small. In the FASTQ data sets, where
original Bonsai uses $85\%$ more space than m-Bonsai (r) while
the advantage is reduced to $23\%$ for webdocs.
Of course, all Bonsai variants are significantly more space-efficient than the TST: on the
FASTQ datasets, by a factor of nearly 60.  Indeed, the TST could not even load
the larger datasets (FASTQ and webdocs) on our machine.



\begin{table}[h]
\centering
\caption{Characteristics of datasets and memory usage (bits per node) for all data structures. TST was not able to complete the process for larger datasets.}
\label{tab:characteristics}
\begin{tabular}{|r|r|r|r|r|r|r|}
\hline 
\multicolumn{0}{|c|}{Datasets}  & \multicolumn{0}{|c|}{$n$} & \multicolumn{0}{|c|}{$\sigma$} & \multicolumn{0}{|c|}{m-Bonsai (r)} & \multicolumn{0}{|c|}{m-Bonsai($\gamma$)} & \multicolumn{0}{|c|}{Bonsai} & \multicolumn{0}{|c|}{TST} \\ 
\hline 
Chess 			& 38610 	 & 75      & 13.99 & 11.94 & 17.51 & 389.56 \\ 
Accidents 		& 4242318 	 & 442     & 16.45 & 14.42 & 19.99 & 388.01 \\ 
Pumsb 			& 1125375 	 & 1734    & 18.95 & 16.93 & 22.51 & 387.52 \\ 
Retail 			& 1125376 	 & 8919    & 22.71 & 20.69 & 26.25 & 384.91 \\ 
Webdocs8 		& 63985704 	 & 364	   & 16.45 & 14.44 & 19.99 & 386.75 \\ 
Webdocs 		& 231232676	 & 59717   & 25.20 & 23.19 & 28.72 & ---\\ 
\hline 
\hline
SRR034939 		& 3095560      & 5     & 8.94  & 6.78 & 12.51 & 385.88 \\ 
SRR034944       & 21005059     & 5     & 8.93  & 6.74 & 12.51 & 385.76 \\ 
SRR034940       & 1556235309   & 5     & 8.93  & 6.77 & 12.51 & --- \\ 
SRR034945       & 1728553810   & 5     & 8.93  & 6.79 & 12.51 & --- \\ 
\hline 
\end{tabular} 

\label{tab:MemUsage}
\end{table}

\paragraph{Tree construction speed.}
In Table \ref{tab:WalltimeInsert} we show the wall clock time in seconds for the
construction of the tree.  Of the three Bonsai implementations, m-Bonsai (r) is
always the fastest and beats the original Bonsai by 25\% for the bigger datasets and about 40\% for the smaller ones.

We believe m-Bonsai (r) may be faster because Bonsai requires moving
elements in $Q$ and one of the bit-vectors with each insertion.
When inserting a node in m-Bonsai (r) each $D[i]$ location is 
written to once and $D$ and $Q$ are not rearranged after that.
Finally, m-Bonsai ($\gamma$) is an order of magnitude slower than ther other
Bonsai variants. 
It would appear that this is due to the time required
to access and decode concatenated $\gamma$-encodings of a block,
append the value and then encode the whole block back with the new value. 

Comparing to the TST, m-Bonsai (r) was comparably fast even on small datasets which fit
comfortably into main memory and often faster (e.g. m-Bonsai is 30\% faster on webdocs).
The difference appears to be smaller on the FASTQ datasets, which have small alphabets.
However, the TST did not complete loading some of the FASTQ datasets.

\begin{table}[h]
\centering
\small
\caption{The wall clock time in seconds for the construction of the Trie. TST was not able to complete the process for larger datasets.}
\begin{tabular}{|r|r|r|r|r|}
\hline 
\multicolumn{0}{|c|}{Datasets}  & \multicolumn{0}{|c|}{m-Bonsai (r)} & \multicolumn{0}{|c|}{m-Bonsai($\gamma$)} & \multicolumn{0}{|c|}{Bonsai} & \multicolumn{0}{|c|}{TST} \\ 
\hline 
Chess		&  0.02		& 	  0.21 	&	  0.08	&	0.02		\\
Accidents	&  2.18		& 	 21.46  &	  3.01 	&	2.31	\\
Pumsb	 	&  0.43		& 	  5.85  &	  0.69 	&	0.57		\\
Retail	 	&  0.22	 	& 	  2.27  &	  0.25 	&	0.31		\\
Webdocs8	& 26.07 	& 	252.59  &	 32.75 	&  18.25	\\
Webdocs		& 96.38 	& 	869.22  &	130.92 	&	---	\\

\hline
\hline
SRR034939	&   0.61	&	   10.25	&	    0.79	&	0.61		\\
SRR034944	&   5.72	& 	   70.83	&	    7.34	&	4.31	\\
SRR034940	& 730.55	& 	6,192.14	&	  970.81 	&	---	\\
SRR034945 	& 841.87 	&   7,199.11	&	1,106.39 	&	---	\\
\hline 
\end{tabular} 
\label{tab:WalltimeInsert}
\end{table}

\paragraph{Traversal speed.}
In Table~\ref{tab:traversal} we show compare the simple linear-time and naive traversals,
using m-Bonsai (r) as the underlying Bonsai representation.
After we construct the trees for the given datasets, we traverse and measure
the performance of the two approaches in seconds. The simple linear-time traversal includes both the preparation and the traversal phase. Since the naive traversal takes
$O(n \sigma)$ time and the simple traversal takes $O(n)$ time, one would expect
the difference to be greater for large $\sigma$. For example, Retail is a relatively small dataset with large $\sigma$, the difference in speed is nearly two orders of magnitude. Whereas all FASTQ datasets are consistently only $2$ times slower.

Surprisingly, even for datasets with small $\sigma$ like the FASTQ, naive approach does worse than the simple linear-time approach.  This may be because the simple linear-time traversal makes fewer
random memory accesses (approximately $3n$) during traversal, while the naive approach
makes $n \sigma = 5n$ random memory accesses for this datasets.  In addition, note that
most searches in the hash table for the naive traversal are \emph{unsuccessful} searches,
which are slower than successful searches.

\begin{table}[h]\small
\centering
\caption{The wall clock time in seconds for traversing the tree using simple and naive approach.}
\begin{tabular}{|r|r|r|r|r|r|r|}
\hline 
\multicolumn{0}{|c|}{Datasets}  & \multicolumn{0}{|c|}{Simple traversal} & \multicolumn{0}{|c|}{Naive traversal} \\ 
\hline 
Chess 			& 0.02 	 	& 0.38		\\ 
Accidents 		& 4.82 	 	& 228.92    \\ 
Pumsb 			& 1.01 	 	& 233.11   	\\ 
Retail 			& 1.04 	 	& 788.36    \\ 
Webdocs8 		& 104.92 	& 6,617.39	\\ 
Webdocs 		& 150.81	& ---    	\\ 
\hline 
\hline
SRR034939 		& 2.61      & 4.52     \\ 
SRR034944       & 24.78     & 41.94     \\ 
SRR034940       & 3,352.81   & 7,662.37    \\ 
SRR034945       & 4,216.82   & 8,026.94    \\ 
\hline 
\end{tabular} 

\label{tab:traversal}
\end{table}

\paragraph{Successful search speed.}
Now we explain the experiment for the runtime speed for successful search operations.
For this experiment we designed our own \emph{search}-datasets, where we randomly picked $10\%$ of the strings from each dataset, shown in Table \ref{tab:WalltimeSearch}. 
After the tree construction, we measured the time needed in nanoseconds 
per successful search operation. 
It is obvious that TST is the fastest approach.
However, m-Bonsai (recursive) remains competitive with TST and consistently faster than Bonsai by at least 1.5 times, whereas m-Bonsai ($\gamma$) in the slowest. 
Note that there is an increase in runtime speed per search operation for all Bonsai data structures as the datasets get bigger, since there are more cache misses.
For TST, we see that Retail with high $\sigma$ is affecting the runtime speed, as TST can search for a child of a node in $O(\log\sigma)$ time.

\begin{table}[h]
\centering
\caption{The wall clock time in nanoseconds per successful search operations.}
\begin{tabular}{|r|r|r|r|r|}
\hline 
\multicolumn{0}{|c|}{Datasets}  & 
\multicolumn{0}{|c|}{m-Bonsai (r)} & 
\multicolumn{0}{|c|}{m-Bonsai($\gamma$)} & 
\multicolumn{0}{|c|}{Bonsai} & 
\multicolumn{0}{|c|}{TST} \\ [1.1ex]
\hline 

Chess		& 130 		& 	1240  &	288	&	59	\\
Accidents	& 187 		& 	1342  &	399	&	60	\\
Pumsb	 	& 134 		& 	1204  &	301	&	55	\\
Retail	 	& 407 		& 	1244  &	418	&	102	\\
Webdocs8	& 296 		& 	1573  &	586	&	61	\\
Webdocs		& 352	 	& 	1705  &	795	&	---	\\

\hline
\hline
SRR034939	& 173	&	1472	&	350	&	65	\\
SRR034944	& 247	& 	1682	&	498	&	66	\\
SRR034940	& 451	& 	1946	&	709	&	---	\\
SRR034945 	& 511 	&   1953	&	718 &	---	\\
\hline 
\end{tabular} 

\label{tab:WalltimeSearch}
\end{table}

`

%% file: conclusion.tex
We have demonstrated a new variant of the Bonsai approach to
store large tries in a very space-efficient manner.  
Not only have we (re)-confirmed that the original Bonsai approach is
very fast and space-efficient on modern architectures,
both m-Bonsai variants we propose
are significantly smaller (both asymptotically and in practice) and
and one of them is a bit faster than the original Bonsai.  
In the near future we intend to investigate other variants, 
to give a less ad-hoc approach to m-Bonsai (recursive),
and to compare with other trie implementations.

Neither of our approaches is very close to the
information-theoretic lower bound of 
$(\sigma \log \sigma - (\sigma-1)\log (\sigma -1))n - O(\log(kn))$ 
bits \cite{DBLP:journals/algorithmica/BenoitDMRRR05}.
For example, for $\sigma = 5$, the lower bound is $3.61n$ bits, while
m-Bonsai ($\gamma$) takes $\sim 5.6M \sim 7n$ bits.  Closing this gap
would be an interesting future direction.  Another interesting
open question is to obtain a practical compact dynamic trie that has a 
wider range of operations, e.g. being able to navigate
directly to the sibling of a node.

\paragraph{Acknowledgements.}
We thank Gonzalo Navarro for helpful comments and in particular suggesting
an asymptotic analysis of the $D$-array representation used in
m-Bonsai (r).